# Enhancing hole mobility in III-V semiconductors

Aneesh Nainani[*], Brian. R. Bennett[#], J. Brad Boos[#], Mario G. Ancona[#], Krishna C. Saraswat

Center for Integrated Systems, Department of Electrical Engineering, Stanford University, Stanford, CA 94305 USA

[#]Naval Research Laboratory, Washington, DC 20375 USA

Transistors based on III-V semiconductor materials have been used for a variety of analog and high frequency applications driven by the high electron mobilities in III-V materials. On the other hand, the hole mobility in III-V materials has always lagged compared to group-IV semiconductors such as silicon and germanium. In this paper we explore the used of strain and heterostructure design guided by bandstructure modeling to enhance the hole mobility in III-V materials. Parameters such as strain, valence band offset, effective masses and splitting between the light and heavy hole bands that are important for optimizing hole transport are measured quantitatively using various experimental techniques. A peak Hall mobility for the holes of 960cm$^2$/Vs is demonstrated and the high hole mobility is maintained even at high sheet charge.

---

[*] email : nainani@stanford.edu

# I. INTRODUCTION

III-V semiconductors have been used extensively for making high electron mobility transistors (HEMTs) and heterojunction bipolar transistors (HBTs) for analog, digital, and mixed-signal high-frequency applications. Recently, with the development of high quality dielectrics on III-V surfaces, III-V materials are also being actively investigated for replacing silicon as the channel material for making MOSFETs for low power / high performance applications[1-3]. Most of these developments have been driven by the promise of higher electron mobilities and the potential benefits of heterostructure engineering with nearly lattice-matched semiconductors[4]. For p-channel devices the advantages have been less clear because the hole mobility in III-V materials has always lagged compared to group-IV semiconductors such as silicon and germanium. In particular, this has been a limiting factor for reducing base resistance in the p-type base of a III-V HBT and for obtaining a high performance channel for III-V p-channel FET.

In the case of silicon, the hole mobility has been greatly enhanced through the use of strain. Up to two times enhancement in hole mobility and drive current is obtained in state-of-the-art p-channel silicon transistors using of uniaxial strain[5]. Theoretical calculations have predicted greater than four times enhancement in the future[5]. III-V systems offer great flexibility for strain engineering via control of the lattice mismatch between the channel and the underlying layers. Furthermore, heterostructure design of III-V transistors can be used to create a quantum well channel for the holes that can be used to further improve the hole transport. Given the lattitude III-V materials offer in material options and strain configurations, a useful first step is to use modeling to help engineer the stack and strain for optimum hole transport.

In 1983 Osburn[6] proposed the use of a strained $In_xGa_{1-x}As$ layer to enhance the hole mobility and thereby improve the performance of the p-channel field effect transistor. Laikhtman et al.[7] presented a modeling study of the InGaAs/AlGaAs system with the objective of discussing the critical parameters influencing hole transport. More recently, Bennett et al. demonstrated the use of biaxial strain in $In_xGa_{1-x}Sb$[8] & $GaSb$[9] channels to enhance the hole mobility and obtained hole mobilities of greater than 1000cm$^2$/Vs at a sheet charge of $1\times10^{12}$/cm$^2$. The key concept behind these schemes for enhancing hole mobility is demonstrated in Figure 1, where a narrow bandgap material is inserted between the wide bandgap layers creating a quantum well for the confined holes. The amount of strain present in the narrow bandgap channel can be varied by engineering the lattice mismatch between the channel and barrier layers. The presence of strain and confinement splits the degeneracy between the light hole (*lh*) and heavy hole (*hh*) bands (Figure 1). This improves mobility both by increasing the occupancy of carriers in the light hole band that has a lower transport effective mass and by reducing the number of final states available for interband scattering of the holes. Note that we are labeling the lowest occupied hole band as the light hole band (Figure 1) which is indeed the band with lighter hole effective mass in our case as discussed later.

The important parameters for designing the quantum well (Figure 1) to enhance hole mobility are the percentage of strain induced in the channel, the valence band offset (VBO) between the narrow bandgap channel and the wide bandgap buffer which determines the maximum number of holes confined in the quantum well, the number and effective mass (m*) of the carriers in the light hole (*lh*) and heavy hole (*hh*) bands, and the energy split between the *lh* and *hh* bands ($\Delta_{lh-hh}$). Also it is important to identify the dominant scattering mechanism, which limits the mobility of this two-dimensional hole gas (2DHG).

In this paper, we first used 8 band k.p bandstructure calculations to identify the optimum channel and strain configuration for high hole mobility in III-V semiconductors. Based on the insight gained from the modeling, two different heterostructure designs with strained antimony (Sb) based channels were fabricated and analyzed. The amount of strain present in the channel was quantified using x-ray diffraction (XRD) analysis. X-ray photoemission spectroscopy (XPS) analysis was used to estimate the valence band offset. Temperature-dependent Hall measurements were performed to identify the dominant scattering mechanisms and mobility spectrum analysis (MSA) was used to estimate the number of carriers in the light and heavy hole bands. Effective mass and the splitting between the light and heavy hole bands were quantified using Shubninov-de Haas oscillations observed in these 2DHG at low temperatures and high magnetic field. Finally, using gated Hall measurements, the hole mobility was measured as a function of the sheet charge in $In_xGa_{1-x}Sb$ channels optimized for hole transport. A hole mobility of 960cm$^2$/Vs at a sheet charge of $1\times10^{12}$/cm$^2$ was measured. The mobility remained more than 3 times higher in comparison with uniaxially-strained silicon even at a high sheet charge of $7\times10^{12}$/cm$^2$.

The rest of the paper is organized as follows. In Section II, we present modeling analysis to predict the optimum material and strain configuration for obtaining high hole mobility. Section III discusses the two heterostructure designs that were fabricated and analyzed for hole transport. Section IV details the experiments that were performed on these stacks to measure the amount of strain, the effective masses and the splitting of the light and heavy hole bands, the valence band offset, and the identification of the dominant scattering mechanism. Section V reports on gated Hall bar measurements and mobility results. Finally, the conclusions are summarized in Section VI. These results demonstrate that high hole mobility can be obtained in III-V materials with prudent material selection and optimum strain and heterostructure design. This will serve as an important step in the development of high performance III-V pMOSFET and III-V npn HBT's. The process of material selection and device design guided by modeling and direct measurement of quantum mechanical parameters that determined the transport can be applied to other material systems as well and should be useful for a broader audience.

## II. MODELING

As done in Ref. 10[10], we used an 8 band k.p approach to model the bandstructure for technologically-relevant arsenic (As) and antimony (Sb) based III-V compounds. An 8x8 Hamiltonian[11] with spin orbit coupling was used to evaluate strain effects under both biaxal and uniaxial conditions. Parameters used for the simulation were calibrated against the bulk bandstructure obtained using the non-local empirical pseudopotential method with and without strain. A self-consistent method was used to calculate the valence subband structure[12]. The use of the 8 band model[13], which includes the interaction between the conduction, light, heavy and split-off hole bands, matches the results from empirical pseudopotential method better than the 6-band k.p model[14] which only accounts for coupling between the light, heavy and split-off bands. This coupling of the conduction band with valence bands is especially important for low bandgap III-V materials such as InAs and InSb. Bandstructures of tertiary $In_xGa_{1-x}As$ and $In_xGa_{1-x}Sb$ were calculated using the band parameters of binary end points using Bowdin's interpolation[15].

Figure 2 plots the isoenergy surfaces of the upper valence band at 2, 25 and 50 meV for silicon, GaAs and InSb. We observe that for GaAs and InSb, the valence-band isoenergy surface remain relatively isotropic with increasing energy as compared to silicon. In the case of silicon, the valence bands are warped to start with and the warping reflective of the non-parabolicity increases with increasing energy (Figure 2). Figure 3 shows the isoenergy surfaces for the upper valence band of GaAs with biaxial and uniaxial-[110] compression. For biaxial stress, the energy surface is an ellipsoid with the energy contours in the x-y plane being circles (Figure 3(a)); the band is *lh*-like in the plane of the stress and *hh*-like out-of-plane. Under uniaxial compression (along [110]), the energy contour for the top band in the x-y plane (Figure 3(b)) is an ellipse with the major axis along [-110] and the minor axis along [110]. Though qualitatively the effect of strain on III-V materials is similar to group-IV elements, the prime differences arise from: (a) the initial isotropy of the valence bands in III-V's as compared to Si/Ge (Figure 2), (b) the lower modulus of elasticity resulting in more strain for the same amount of stress in III-V's as compared to Si/Ge (Table. I) and (c) the additional effect due to increased mixing with the conduction band especially for III-V's with low-bandgap.

Table. I lists the low field hole mobility for various III-V binaries. It is well known that for electrons in III-V semiconductors, polar scattering is the dominant scattering mechanism as non-polar optical phonons do not interact with the electrons due to the s-like spherical symmetry of the conduction band in III-V's[16]. In the case of holes, however, both deformation potential and polar scattering mechanisms are important for mobility calculations. In Table I, we observe that varying the group III element while keeping the group V element the same (i.e. InAs → GaAs) does not change the hole mobility appreciably while a large change in hole mobility is observed when the group V element is varied (i.e. InP → InAs → InSb). This is a consequence of the fact that the valence bands in III-V materials primarily derive from the p-orbitals of the anion[17]. Figure 4 plots the low field hole mobility of $In_xGa_{1-x}As$ and $In_xGa_{1-x}Sb$ at a sheet charge density of $1\times10^{12}/cm^2$. We accounting for acoustic/optical deformation potential, alloy and polar scattering mechanisms to calculate the hole mobility[12, 13]. We observe that the hole mobility for the tertiary III-Vs roughly stays in between those of the binary end points. Thus antimonides have significantly higher hole mobilities than arsenides. Note that at higher fields the mobility of the ternaries is expected to be less than that of the end point binaries due to increased alloy scattering.

Next we study the effectiveness of strain in enhancing the hole mobility in III-V semiconductors. Figure 5 plots the hole mobility enhancement for (001) substrate orientation and a fixed 2% strain. The figure is in the form of a polar plot where the distance from the origin represents the enhancement in mobility and the direction represents the channel direction[18]. Figure 5(a) plots the enhancement for 2% biaxial compression/tension, while Figure 5(b) is for fixed 2% uniaxial compression/tension with the uniaxial strain always applied along the channel direction. Results for compression/tension are represented with solid/hollow circles, respectively. We make the following observations: (a) compressive strain is always better than tensile strain for hole mobility enhancement in III-V's. (b) Enhancement with biaxial strain is approximately isotropic while the strain response with uniaxial strain has a large directional dependence with the [011] channel direction being optimal. This can be related to Figure 3 where the isoenergy surface is isotropic in the transport plane for biaxial strain while highly anisotropic for uniaxial strain. (c) We can get up to 4.3 and 2.3 times enhancement with 2% uniaxial and biaxial compression (Figure 5), respectively. Figure 6 plots the results for varying levels of biaxial strain that can be introduced by engineering the lattice mismatch between the channel and the buffer layers during the growth of these materials. We note that the antimonides have twice as high an unstrained hole mobility in comparison to arsenides which can be enhanced further with the use of biaxial compression. Thus compressively-strained antimonide channels are the most promising candidates to obtain high hole mobility in III-V materials.

## III. HETEROSTRUCTURE DESIGN

Figure 7 plots the band energy of different antimonide-containing semiconductors versus their lattice constants. Looking at Figure 7, we investigated two approaches for achieving biaxial compression in Sb-based channels while having sufficient offset in the valence band for confining the 2DHG. Approach (A) uses $In_xGa_{1-x}Sb$ channel and $Al_yGa_{1-y}Sb$ barrier. Approach (B) uses a binary GaSb channel with an $AlAs_xSb_{1-x}$ barrier. In both cases the lattice constant of the channel material ($In_xGa_{1-x}Sb$ for Approach A and GaSb for Approach B) is made slightly higher than the lattice constant of the barrier layer so as to introduce compressive strain into the channel.

We fabricated and studied heterostructures based on both of these approaches. Figure 8 shows the different layers for the heterostructures for the two approaches that were grown using molecular beam epitaxy (MBE) on semi-insulating GaAs (100) substrates. A micron-thick buffer layer was employed to absorb the lattice mismatch between the GaAs substrate and the channel layers. The $AlAs_xSb_{1-x}$ buffer for approach B is grown as a superlattice of AlAs and AlSb[9], to allow better control of the composition compared to a random alloy of $AlAs_xSb_{1-x}$. A 1µm thick buffer layer of $AlAs_xSb_{1-x}$ consisted of 666 periods of the AlAs / AlSb superlattice layers. Modulation doping was utilized using Be-doped layers either below or above the channel (Figure 8). The use of modulation doping allows the desired carrier concentration to be achieved while reducing the Columbic scattering due to dopants. More details on the growth are given elsewhere[8, 9]. Cross-sectional transmission electron microscope (TEM) analysis was performed on these samples. Figure 9 shows a cross-sectional image of the buffer layer. The threading dislocations and misfit defects at the interface that arise from the lattice mismatch between the substrate and buffer are marked in the figure. We observed using TEM analysis that most of the dislocations/defects are contained in the buffer layer and it was possible to obtain good crystal quality near the channel layers (Figure 8).

## IV. EXPERIMENTS

In this section, we present various experiments conducted on the heterostructures discussed in the previous section to quantify the amount of strain present in the channel, the VBO, the effective masses and the number of carrier in the light and heavy hole bands, splitting between the light/heavy bands, and the mobility obtained. Table. II outlines the experiments performed and the corresponding figures. The different samples studied are listed in Table. III.

### A. Strain

Compressive strain was introduced intentionally in the channel of our stacks to enhance the hole mobility. We used high resolution XRD analysis to quantify the strain present in the channel and also to check for any residual strain present in the metamorphic buffer as it absorbs the lattice mismatch with the GaAs substrate. Figure 10 shows the rocking curves near the (004) GaAs peak for sample A1 (with $In_{0.41}Ga_{0.59}Sb$ channel) and sample B1 [GaSb channel with superlattice of $(AlAs)_x(AlSb)_{1-x}$]. The different peaks in the rocking curve for sample A1 (Figure 10(a)) correspond to the peak from the GaAs substrate, the metamorphic AlGaSb buffer and the peak from the InGaSb channel. For sample B1, we see the main and satellite peaks from the $(AlAs)_xAlSb_{1-x}$ which are characteristics of the digital superlattice (n=-1,0,and+1) that was used in the metamorphic buffer. The GaSb channel peak gets buried in the n = 0 satellite peak from the superlattice. The thicknesses for the AlSb and AlAs layers are determined by matching the experimental results with simulations and the $AlAs_xSb_{1-x}$ ternary composition is calculated using Vegard's law. Table III summarizes the results. Reciprocal lattice scans were performed on a few samples to further quantify the strain present and to check for the presence of residual strain in the metamorphic buffer layer. Figure 11 plots the scan around the (004) and (115) reciprocal lattice points. The peaks from the metamorphic buffer showed that the buffer was 97% relaxed and the channel was pseudomorphically strained with respect to the buffer layer. The epilayer peaks were broadened as a result of a high density of misfit dislocations generated by the large 7-8% lattice mismatch with the substrate.

### B. Valence band offset measurement

A sjgnificant valence band offset (VBO) is needed in order to confine the 2DHG in the channel layer. An experimental measurement of the offset is important as it might be affected by the quantization and strain in the narrow quantum well channel. We used low energy XPS analysis which is the most accurate method to measure the valence band offset[19]. A timed etch was used to etch various layers in the heterostructure shown in Figure 8. VBO was calculated by taking the difference between the valence band spectrum from the channel and barrier layers (Figure 12 and Figure 13). The signal from the channel versus the buffer layer is differentiated by monitoring the element that is different between the channel and buffer layer, i.e., indium in the InGaSb/AlGaSb stack in approach A (Figure 12) and gallium in the GaSb/AlAs$_y$Sb$_{1-y}$ channel in approach B (Figure 13). The valence band offset is estimated to be ~0.3eV for Sample A1 (Figure 12) and 0.6eV for Sample B1 (Figure 13). Approximately twice as high a VBO is achieved using the AlAs$_x$Sb$_{1-x}$ barrier (approach B) as compared to Al$_y$Ga$_{1-y}$Sb barrier (approach A), which was also expected from the bulk band lineup in Figure 7. Both band offsets are sufficient to confine the 2DHG, and are comparable or higher then the corresponding numbers in the Si$_{1-x}$Ge$_x$/Si system[20, 21].

## C. Transport measurements

Since the 2DHG in our structures is confined in a strained quantum well, it is important to quantify the mobility in our samples and identify the scattering mechanisms that are limiting the mobility. It is especially important to check whether the mobility is limited by phonon-related effects that are inherent to the semiconductor, or by scattering at the interface. Fortunately, the interface and phonon related scattering mechanisms have different temperature dependences, thus allowing them to be dstinguished. Temperature-dependent Hall measurements were performed on the samples with the temperature varying from 2K-300K. Figure 14 plots the hole mobility ($\mu_h$) and and sheet charge (N$_S$) values obtained from these measurements. For samples A1, A2 and B1, a T$^{-3/2}$ temperature dependence of hole mobility, characteristic of a mobility limited by phonon scatterning, is seen in the temperature range of 150-300K[16, 22]. Also the sheet charge remains nearly constant as a function of temperature in these samples as expected from their use of modulation doping. For samples A3 and B2, strain relaxation occurs due to either the channel thickness exceeding the critical limit (A3) or the lattice mismatch exceeding the maximum possible at that particular thickness (B2). We observe that the temperature dependence weakens to T$^{-1}$, an effect that suggests a mobility limited by interface defects[16]. Also there is a significant drop in the low temperature mobility in these samples, which is more sensitive to interface quality. Samples A3 and B2 also exhibit a slight freezeout of sheet charge at low temperature, again suggesting a poor interface and/or the presence of dislocations as a result of strain relaxation in these samples. Thus it appears the room temperature mobility in our samples is limited by phonon-based scattering mechanisms when the strain in confined in the channel, and limited by interface-related scattering when there is a relaxation of the strain. We measured a maximum hole mobility of 960cm$^2$/Vs at a sheet charge of ~1×10$^{12}$cm$^2$/Vs. To further confirm our interpretation, a high temperature anneal (600°C/60s) was performed on one of the samples during MBE just before the channel growth. This is expected to improve the quality of the channel/barrier interface, and indeed we observed a higher low-temperature hole mobility (Figure 15), but only a slight increase in the hole mobility at room temperature, which again is limited by phonon scattering.

## D. Mobility spectrum analysis

Mobility spectrum analysis is a valuable technique that allows one to check for parallel conduction in a heterostructure with different layers and to quantify the number of carrier in each band[23]. In this technique, the magnetic field is varied from 0-9T to check for the presence of multiple conductive channels. The Hall coefficient and the resistivity are measured as a function of the varying magnetic field for various temperatures. The conductivity tensors $\sigma_{xx}$ and $\sigma_{xy}$ can be extracted from the measured Hall coefficient ($R_H$) and resistivity using the equations below[24]:

$$E_x = \rho_{xx} J_x \tag{1}$$

$$\rho_{xx} = \frac{\sigma_{yy}}{\sigma_{xx}\sigma_{yy}+\sigma_{xy}^2} \quad , \quad \rho_{xy} = -\frac{\sigma_{xy}}{\sigma_{xx}\sigma_{yy}+\sigma_{xy}^2} \tag{2}$$

$$E_y = \rho_{yx} J_x = R_H B J_x \quad , \quad R_H = \frac{\rho_{yx}}{B} = \frac{\sigma_{xy}}{\sigma_{xx}\sigma_{yy}+\sigma_{xy}^2}\frac{1}{B} \tag{3}$$

The components of the conductivity tensor are related to the charge and mobility of individual carriers in a N-carrier system as:

$$\sigma_{xx} = \sum_1^N \frac{n_i q_i \mu_i}{1+\mu_i^2 B^2} \quad , \quad \sigma_{xy} = \sum_1^N \frac{n_i q_i \mu_i^2 B}{1+\mu_i^2 B^2} \tag{4}$$

where, $n_i$ and $\mu_i$ are the sheet charge and mobility corresponding to the i[th] carrier, respectively. Figure 16 (a) and Figure 16(b) plot the components of the conductivity tensor as a function of the magnetic field measured at many temperatures. A least squares algorithm is used to fit the experimental data assuming one, two and three types of carriers as per equation (4). We find that the fit to the experimentally measured conductivity tensors (Figure 16) is excellent when we assume that only the *lh* and *hh* bands are occupied, thus ruling out any parallel conduction in the heterostructure stack. The number and mobility of the carriers in the *lh* and *hh* bands were calculated as per equation (4) and plotted in Figure 17 for different temperatures. We observe for sample A1 (with a sheet charge of ~$1 \times 10^{12}$/cm$^2$) that only the *lh* band is occupied at 2K (Figure 17), that the *hh* band starts to become occupied as the temperature increases, and that the number of carriers in *lh* and *hh* bands becomes comparable at room temperature.

## *E. Effective mass and the splitting between the light / heavy hole bands*

Shubnikov-de-Haas (SdH) oscillations were observed in our samples at low temperature (2-20K) and high magnetic fields (0-9 Tesla). Figure 18(inset) plots the scan of sheet resistance with magnetic field for sample A1 with a mobility of 4500cm$^2$/Vs and a sheet charge of $1.3 \times 10^{12}$/cm$^2$ at 2K. An oscillatory behavior can clearly be seen superimposed on a parabolic dependence with magnetic field. The oscillatory behavior is plotted versus magnetic field in Figure 18 for various temperatures, with the parabolic dependence due to hole-hole interactions[25] removed. It is well known that these oscillations occur only when there is a low density of defects in the bulk of the channel or at its interfaces; hence observation of SdH oscillations provides more evidence of the good crystal quality in the channel and at the interfaces. Moreover, no oscillations were observed in the samples A3 & B2 where relaxation of strain occurred leading presumably to a deterioration of interface quality.

Besides confirming good crystal quality, significant quantitative information such as the effective masses of carriers in the light and heavy hole bands (m*) and the splitting between the light and heavy hole bands ($\Delta_{lh-hh}$) can be extracted from SdH oscillations. Figure 19 plots the oscillatory behavior at 2K versus the inverse of the magnetic field (1/B) at a sheet charge of $1.1\times10^{12}$/cm$^2$. The oscillations appear periodic in nature and a Fast Fourier Transform (FFT) of the data shows that the oscillations are harmonic with a single frequency (Figure 19(inset)) implying occupancy of only the lowest energy light hole subband at this sheet charge density. The effective mass of the light hole band is calculated from the temperature dependence of the SdH oscillations. The peak amplitude of the oscillations ($\Delta\rho_p$) can be described using the Ando formula[26] as:

$$\frac{\Delta\rho}{\rho_0} = R_s V \frac{\xi}{\sinh \xi} exp\left(\frac{-\pi}{\omega_c \tau_q}\right) \quad (5)$$

where, $\rho_0$ is the resistance at zero B, $\tau_q$ is the quantum lifetime, $\xi = 2\pi^2 kT/\hbar\omega_c$ and $\omega_c = eB/m^*$. The prefactor $R_s$, is associated with Zeeman splitting while V is usually set equal to 4. $R_S$ and V are assumed to be independent of magnetic field and temperature in which case they were not involved in the following analysis. $\ln\left(\frac{\Delta\rho_p}{\rho_0}\right)$ is plotted versus $\ln\left(\frac{\xi}{\sinh \xi}\right)$ in the temperature range of 2-10K, for various values of B at which the peak occurs. $m_{lh}^*$ was calculated as the value for which a gradient of unity is obtained in each sample. Table III summarizes the $m_{lh}^*$ values for the different samples. The m*$_{lh}$ of 0.094m$_0$ obtained in sample A1 was verified using cyclotron resonance. The presence of strain and confinement in these samples leads to a reduction of hole effective mass to 0.094m$_0$, which is close to the value typically found for the electron effective mass in III-V materials.

Figure 20 plots the oscillatory behavior on the same sample at a sheet charge of $3.5\times10^{12}$/cm$^2$. We see that now there are two frequencies beating, and an FFT of the data correspondingly shows two dominant peaks (Figure 20(inset)) meaning that both the *lh* and *hh* bands are occupied at this sheet charge density. The energy splitting between the *lh* and *hh* bands ($\Delta_{lh-hh}$) can be calculated from the onset at which the second frequency starts to occur and was estimated to be between 70-80meV from sample A1. In the well-studied case of strain in silicon it is known that the splitting between the light and heavy hole bands must be more than the energy of the optical phonon in silicon (60meV) if there is to be effective suppression of interband scattering between the light and heavy hole bands[27]. The optical phonon energy is ~28meV for GaSb and ~30meV for InSb, and thus our measured energy splitting would seem to be sufficient for suppressing the interband scattering and giving a large enhancement in hole mobility. The effective mass of the carriers in the heavy hole bands ($m_{hh}$*) can be calculated from the temperature dependence of the second oscillation, and was found to be $0.3 \pm 0.05$ m$_0$ for sample A1. The larger uncertainty in the measurement of the heavy hole mass arises from the superposition of the two oscillations from the light and heavy hole bands when both of the bands are occupied making it more difficult to separate the individual components.

## V. GATED HALL MEASUREMENTS

Although we obtained a high hole mobility of 960cm$^2$/Vs at a sheet charge of ~ $1\times10^{12}$/cm$^2$Vs, it is the mobility at high sheet charge which determines the $I_{ON}$ for MOSFET applications. For this purpose, gated Hall bar structures were fabricated as shown in Figure 21 using Al$_2$O$_3$ deposited by atomic layer deposition and with a Pt gate[28]. With this test structure, the sheet charge can be varied in the channel and the hole mobility can be measured as a function of sheet charge. Figure 21 plots the result for Sample A1. We observe that not only is the mobility higher in comparison to strained[29, 30] (universal[31]) silicon at low sheet charge, but the superior hole mobility is maintained even at high sheet charge. The hole mobility in our samples is more then 3 times (>4 times) higher than uniaxially strained Si (universal Si) even for a high sheet charge of $7\times10^{12}$/cm$^2$.

## VI. SUMMARY

In summary, we find that the valence bands, and hence the hole mobility of III-V channels, are highly dependent on the choice of group V elements and weakly influenced by the group III element. III-V antimonides have a hole mobility that is twice that found in the III-V arsenides. Bandstructure modeling results obtained by using an 8 band k.p approach indicate that the mobility can be enhanced up to 4.3/2.3 times by imposing a 2% uniaxial/biaxial compression on the channel material.

Heterostructure stacks with a compressively-strained antimonide channel were fabricated using two different approaches and analyzed experimentally. TEM analysis, high hole mobility, and observation of SdH oscillations all confirmed the good crystal quality in the channel region. The valence band offset between the channel and barrier was sufficient to confine the carriers in the channel layer, and no parallel conduction was observed using MSA analysis. The presence of strain and confinement in the channel material reduced the *lh* effective mass to 0.094m$_0$, and split the degeneracy between the lh and hh bands by 70-80meV. These effects lead to a significant enhancement in the hole mobility with a maximum hole mobility of 960cm$^2$/Vs at a sheet charge of ~$1\times10^{12}$/cm$^2$ being achieved.

The high mobility in our stacks was maintained even at high sheet charge density and the mobility was more than 3 times higher in comparison to strained silicon at a high sheet charge of $7\times10^{12}$/cm$^2$. This study demonstrates that a high hole mobility can be obtained in III-V materials and will motivate the development of high performance III-V MOSFETs and improvement of the base resistance in HBTs.

.

## ACKNOWLEDGEMENT

Aneesh Nainani would like to thank Intel Corporation for a PhD fellowship. We would like to thank Shyam Raghunathan, Daniel Witte, Masaharu Kobayashi, Toshifumi Irisawa and Tejas Krishnamohan for their help and useful discussions. This work was partially supported by the Office of Naval Research and Intel Corporation.

TABLE I. Relevant properties of different semiconductor materials at 300K. Note that III-V's (antimonides in particular) have lower elasticity constants than silicon.

|      | $E_g$ (eV) | $m_{lh}$ | $m_{hh}$ | Bulk Mod. (dyne/cm2) | $\mu_h$ (cm2/Vs) |
|------|------|-----------|-----------|-----------|------|
| GaP  | 2.26 | $0.14 m_o$ | $0.79 m_o$ | $8.8 \times 10^{11}$ | 150 |
| GaAs | 1.42 | $0.08 m_o$ | $0.51 m_o$ | $7.1 \times 10^{11}$ | 400 |
| GaSb | 0.72 | $0.05 m_o$ | $0.4 m_o$ | $5.63 \times 10^{11}$ | 800 |
| InP  | 1.34 | $0.09 m_o$ | $0.6 m_o$ | $8.8 \times 10^{11}$ | 200 |
| InAs | 0.35 | $0.03 m_o$ | $0.41 m_o$ | $5.8 \times 10^{11}$ | 450 |
| InSb | 0.17 | $0.02 m_o$ | $0.43 m_o$ | $4.7 \times 10^{11}$ | 850 |
| Si   | 1.1  | $0.17 m_o$ | $0.49 m_o$ | $9.9 \times 10^{11}$ | 450 |
| Ge   | 0.67 | $0.043 m_o$ | $0.33 m_o$ | $7.5 \times 10^{11}$ | 1600 |

TABLE II. Parameters measured, technique used and corresponding figures

| Property Measured | Technique Used | Corresponding Figures |
|---|---|---|
| Crystal quality | HR-TEM | Figure 8-9 |
| Strain | HR-XRD | Figure 10-11 |
| VBO | XPS | Figure 12-13 |
| Transport analysis | Hall measurments | Figure 14-15 |
| Number of holes in *lh/hh* bands | MSA | Figure 16-17 |
| Effective mass | SdH | Figure 18 |
| $\Delta_{lh\text{-}hh}$ | SdH | Figure 19-20 |
| Mobility | Gated Hall | Figure 21 |

TABLE III. Details on the samples studied. Mobility and sheet charge ($N_S$) at 300K measured using Hall measurements are listed along with value of light hole effective

mass measured using Shubnikov–de Haas oscillations [#] For samples A3 and B2 the ideally targeted value of strain is listed. Mobility degradation is observed due to strain relaxation in these samples due to channel width exceeding the critical layer thickness.

| Sample | Channel (Thickness (Å)) | Barrier | Strain (%) | $\mu_{Hall}$ (300K) ($cm^2V^{-1}s^{-1}$) | $N_S$ ($cm^{-2}$) | $m^*$ ($m_0$) |
|---|---|---|---|---|---|---|
| A1 | $In_{0.41}Ga_{0.59}Sb$ (75Å) | $Al_{0.7}Ga_{0.3}Sb$ | 1.8 | 960 | $1.3 \times 10^{12}$ | 0.099 |
| A2 | $In_{0.41}Ga_{0.59}Sb$ (75Å) | AlSb | 1.9 | 900 | $0.94 \times 10^{12}$ | 0.094 |
| A3 | $In_{0.41}Ga_{0.59}Sb$ (125Å) | $Al_{0.7}Ga_{0.3}Sb$ | 1.8[#] | 621 | $1.0 \times 10^{12}$ | Strain Relaxation: No Oscillations |
| B1 | GaSb (75Å) | $AlAs_{0.219}Sb_{0.781}$ | 1.06 | 880 | $1.5 \times 10^{12}$ | 0.12 |
| B2 | GaSb (75Å) | $AlAs_{0.238}Sb_{0.762}$ | 1.48[#] | 600 | $1.27 \times 10^{12}$ | Strain Relaxation: No Oscillations |

FIGURE CAPTIONS

Figure 1: Valence band offset (VBO), amount of strain, efftctve mass (m*) and splitting ($\Delta_{lh-hh}$) between the light hole (*lh*) and heavy hole (*hh*) offsets are important parameters for obtaining high hole mobility in III-V heterostructures.

Figure 2 : Isoenergy surfaces for upper valence band in silicon, GaAs and InSb at 2meV / 25 meV / 50 meV.

Figure 3 : Isoenergy surface (left) and 2D energy contour along the transport plane (right) for upper valence band in GaAs for (a) biaxial compression and (b) uniaxial compression.

Figure 4: Calculated hole mobility for varying stoichiometries in $In_xGa_{1-x}As$ and $In_xGa_{1-x}Sb$. Antimonides have twice high hole mobility compared to arsenides.

Figure 5: Polar plot showing calculated hole mobility enhancement for (a) 2% biaxial strain and (b) 2% uniaxial strain. Hollow/solid symbols represent tension/compression. The substrate orientation was (100) while the angle along the plot represents the different directions along which the channel of the transistor can be oriented. For uniaxial strain the strain was applied parallel to the transport direction.

Figure 6: Calculated mobility enhancement for varying amount of biaxial strain, which can be achieved during MBE growth. Positive values represent biaxial compression while negative strain represents biaxial tension.

Figure 7: Two different approaches for obtaining compressively strain Sb-channel. Approach A uses InGaSb channel and AlGaSb barrier. Approach B utilizes GaSb channel and AlAsSb barrier.

Figure 8 : Cross-section showing the different layers in a quantum-well heterostructure with (A) $In_XGa_{1-X}Sb$ and (B) GaSb channel. The $AlAs_XSb_{1-X}$ layer is composed of AlSb/AlAs short-period superlattice. Also shown are high resolution TEM images around the channel region.

Figure 9: (a) Dislocations and (b) misfit defects in the buffer layer which accommodates the large lattice mismatch between the channel and the GaAs substrate.

Figure 10: High Resolution XRD scans on the samples A1 (top) & B1 (bottom) near (004) GaAs peak. For sample B1 which uses (AlAs)AlSb as the buffer we observe main and satellite peaks characteristic of the digital superlattice.

Figure 11: Reciprocal lattice scan on sample B1 around GaAs (004) and (115).

Figure 12: VBO for sample A1 (approach A1) is calculated by taking the difference in the valence band spectrum from the $In_xGa_{1-x}Sb$ channel and $Al_yGa_{1-y}Sb$ buffer.

Figure 13: VBO for sample B1 (approach B) is calculated by taking the difference in the valence band spectrum from the GaSb channel and $AlAs_ySb_{1-y}$ buffer.

Figure 14: Hole mobility ($\mu_h$) and sheet charge ($N_s$) are measured as a function of temperature using Hall measurements for samples: A1, A2, A3 (top) and B1, B2 (bottom).

Figure 15: A high temperature anneal (600°C/60s) before channel growth to optimize the interface results in a large increase in low temperature mobility but gives only slight gain (900cm$^2$/Vs to 940cm$^2$/Vs) in mobility at 300K.

Figure 16: Conductivity tensors ($\sigma_{xx}$ and $\sigma_{xx}$) are measured as a function of magnetic field (B) for various temperatures. Mobility spectrum analysis (MSA) on the data confirms that there is no parallel conduction in the stack and is used to estimate number of carriers in *lh/hh* bands and their mobility (Figure 17).

Figure 17: (a) Number and (b) mobility of carriers in the *llight* (*lh*) & heavy hole (*hh)* bands as a function of temperature for sample A1.

Figure 18: Shubnikov-de-Haas (SdH) oscillations in sheet resistance (inset) are observed at low temperatures and high magnetic field. Temperature dependence of these oscillations is used to calculate **m*** (Table III).

Figure 19: SdH oscillations at 2K are plotted vs. 1/B for sheet charge of 1.1×10$^{12}$/cm$^2$. The oscillatory behavior is periodic in nature with a single dominant frequency, indicating that only the *lh* band is occupied at this sheet charge.

Figure 20: SdH oscillations at 2K are plotted vs. 1/B for sheet charge of 3.5×10$^{12}$/cm$^2$. The oscillations are combinations of 2 dominant frequencies, indicating that both *lh* and *hh* band is occupied at this sheet charge.

Figure 21: Hole mobiliyt ($\mu_h$) is measured as a function of sheet charge ($N_s$) using gated hall measurements. Reported values in (strained) silicon are also plotted for comparison.

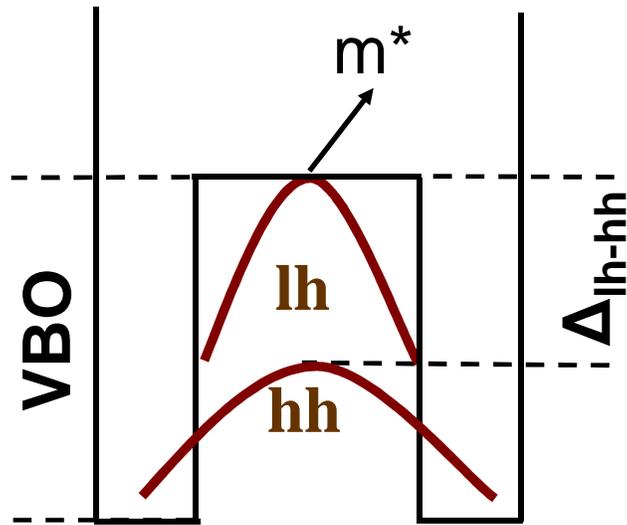

Figure 1 ; Valence band offset (VBO), amount of strain, efftctve mass (m*) and splitting ($\Delta_{lh-hh}$) between the light hole (*lh*) and heavy hole (*hh*) offsets are important parameters for obtaining high hole mobility in III-V heterostructures.

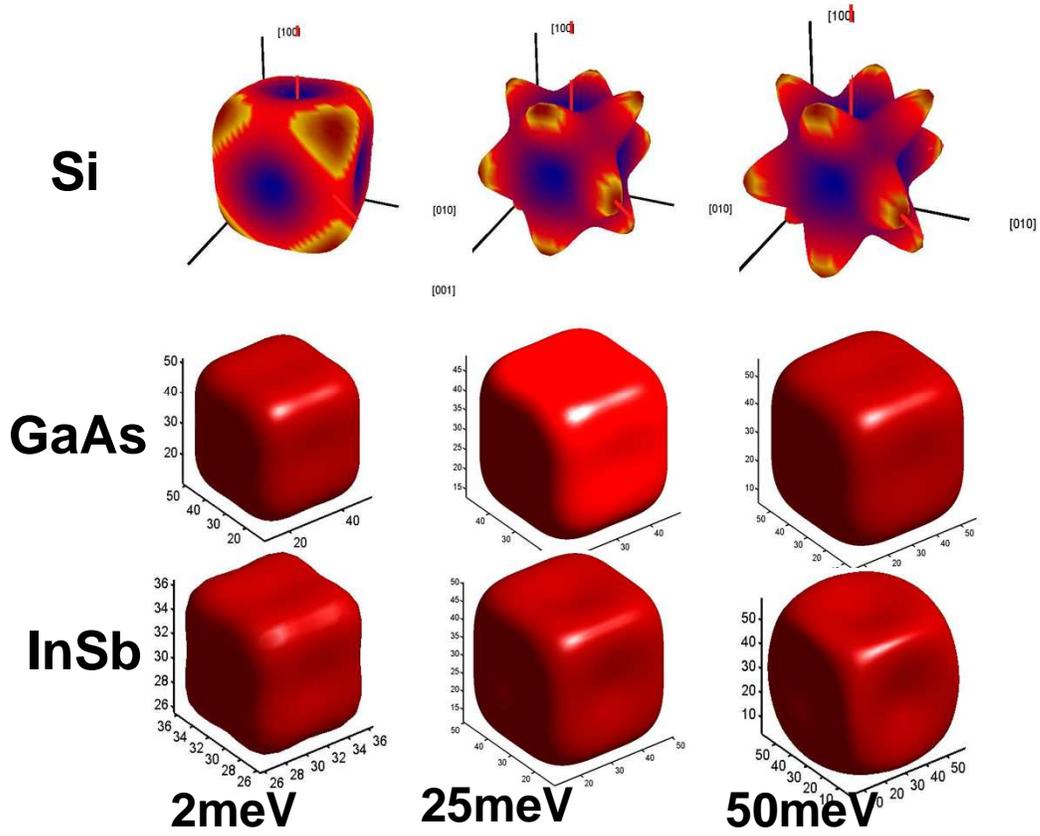

Figure 2 : Isoenergy surfaces for upper valence band in silicon, GaAs and InSb at 2meV / 25 meV / 50 meV.

## (a) Biaxial

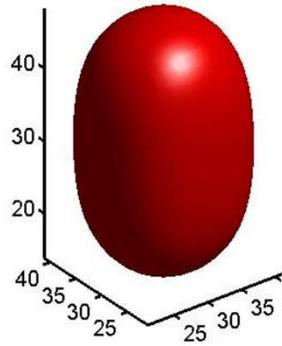 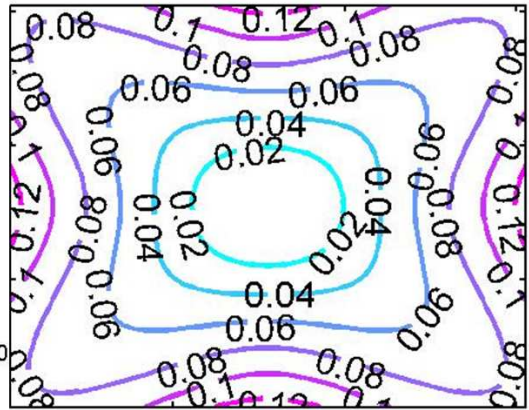

## (b) Uniaxial

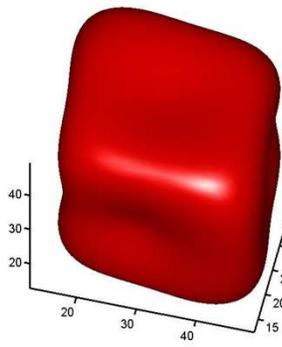 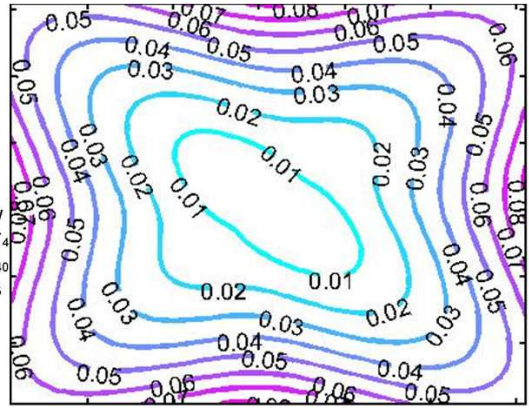

## Upper VB under compression

Figure 3 : Isoenergy surface (left) and 2D energy contour along the transport plane (right) for upper valence band in GaAs for (a) biaxial compression and (b) uniaxial compression.

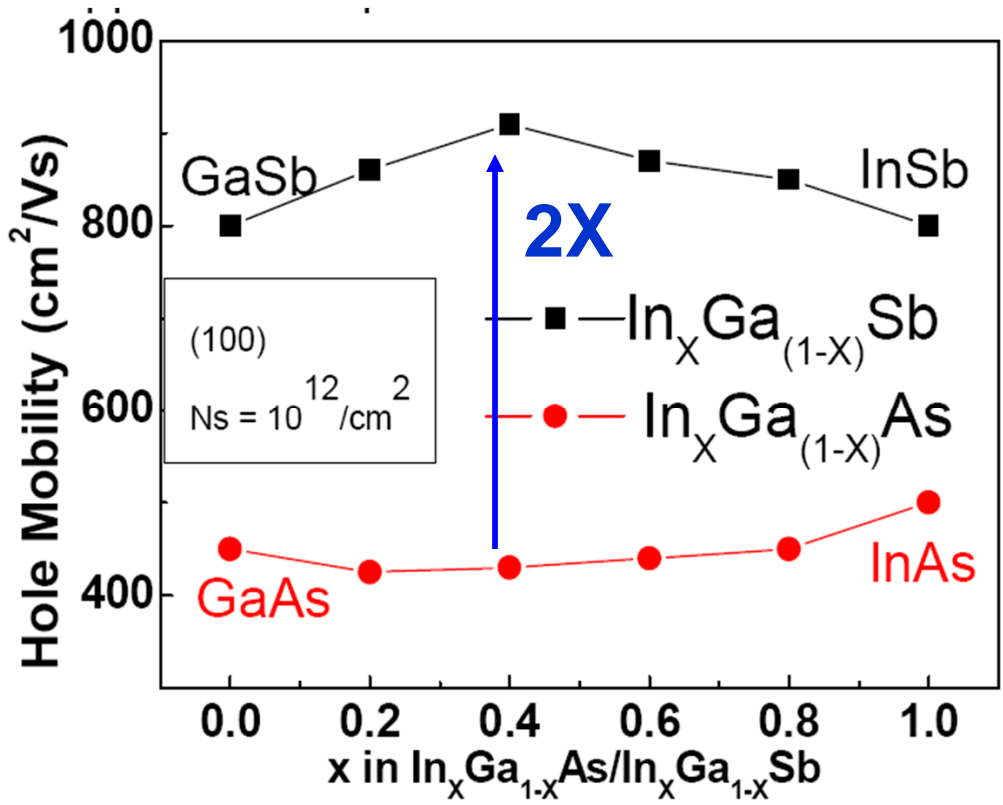

Figure 4: Calculated hole mobility for varying stoichiometries in $In_xGa_{1-x}As$ and $In_xGa_{1-x}Sb$. Antimonides have twice high hole mobility compared to arsenides.

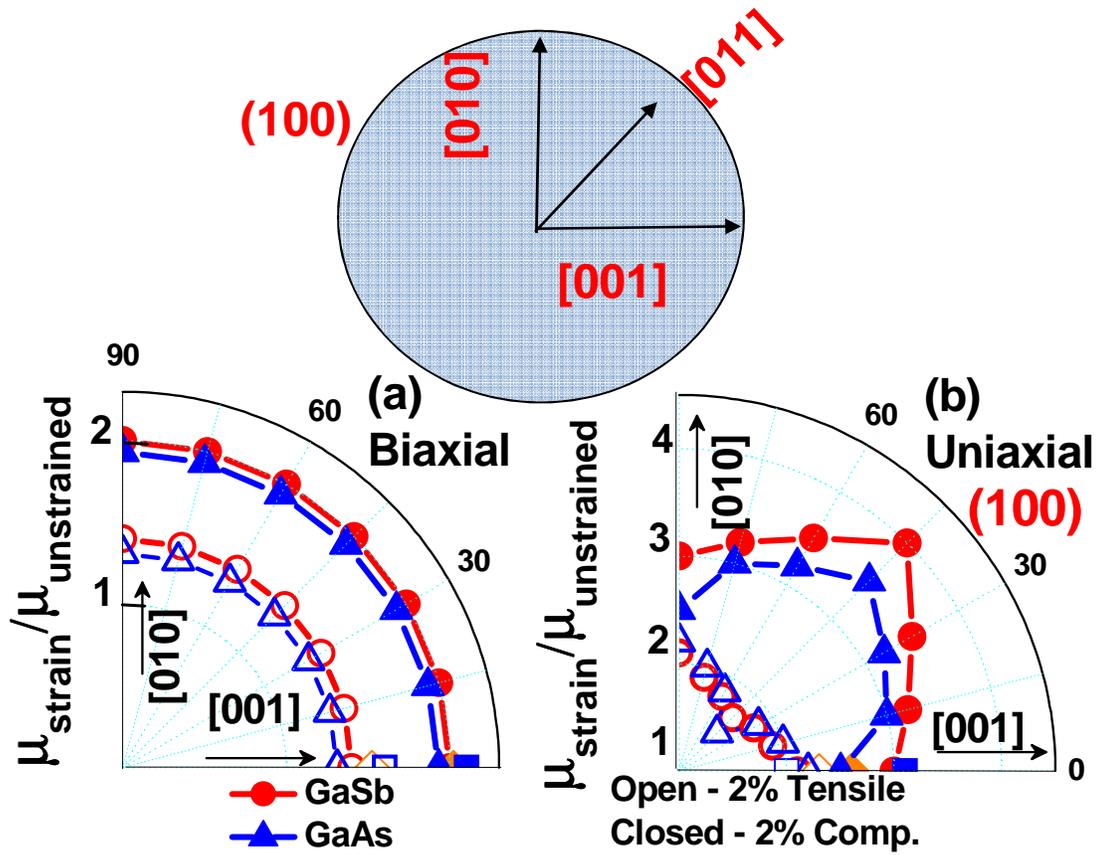

Figure 5: Polar plot showing hole mobility enhancement for (a) 2% biaxial strain and (b) 2% uniaxial strain. Hollow/solid symbols represent tension/compression. The substrate orientation was (100) while the angle along the plot represents the different directions along which the channel of the transistor can be oriented. For uniaxial strain the strain was applied parallel to the transport direction.

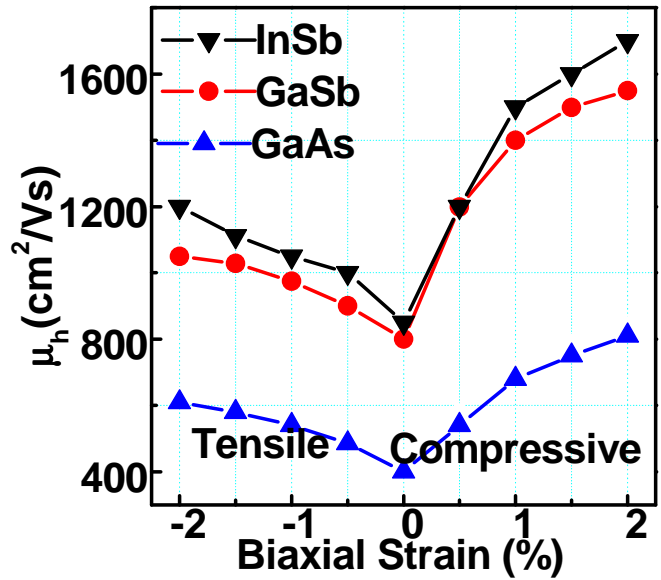

Figure 6: Mobility enhancement for varying amount of biaxial strain, which can be achieved during MBE growth. Positive values represent biaxial compression while negative strain represents biaxial tension.

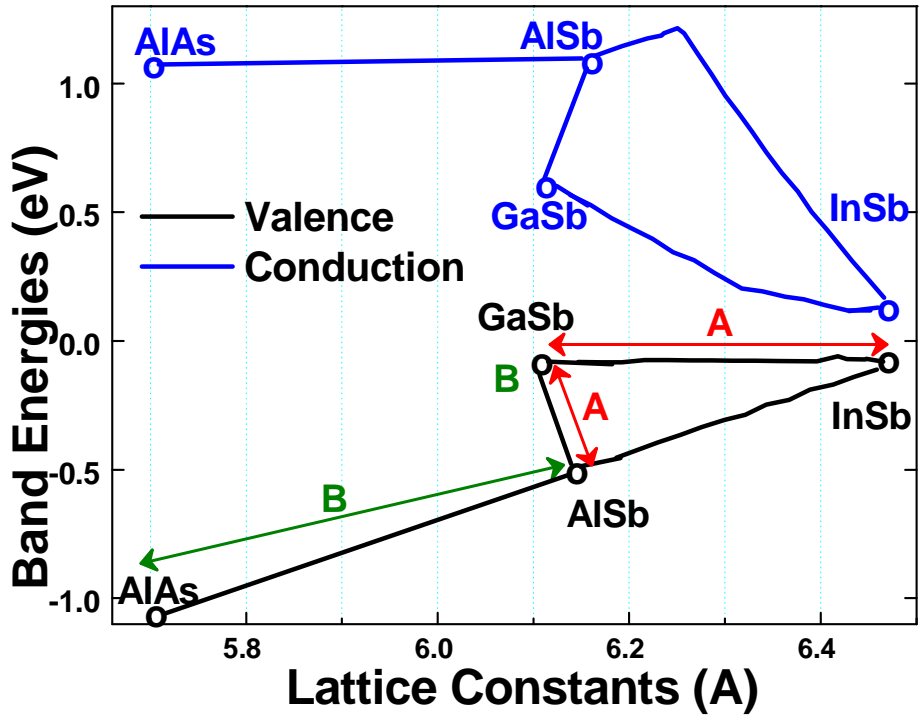
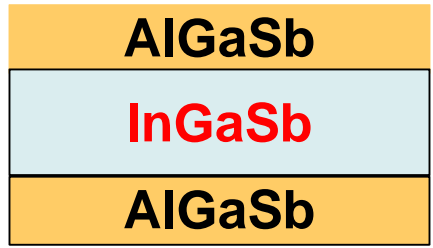
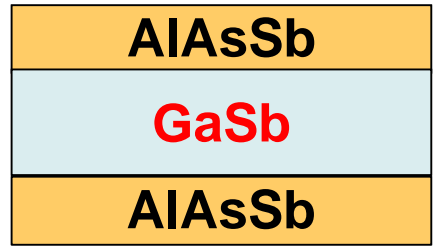

Figure 7: Two different approaches for obtaining compressively strain Sb-channel. Approach A uses InGaSb channel and AlGaSb barrier. Approach B utilizes GaSb channel and AlAsSb barrier.

## Approach A

| InAs 2nm |
| In$_{0.2}$Al$_{0.8}$Sb 4nm |
| Al$_{0.7}$Ga$_{0.3}$Sb (Be) 5nm |
| Al$_{0.7}$Ga$_{0.3}$Sb 21nm |
| In$_{0.41}$Ga$_{0.59}$Sb 7.5nm |
| Al$_{0.7}$Ga$_{0.3}$Sb 1.5µm |
| SI GaAs substrate |

## Approach B

| InAs 2nm |
| In$_{0.2}$Al$_{0.8}$Sb 4nm |
| AlAs$_x$Sb$_{1-x}$ 10.5nm |
| GaSb 7.5/10nm |
| AlAs$_x$Sb$_{1-x}$ 10.5nm |
| AlAs$_x$Sb$_{1-x}$ (Be) 4.5nm |
| AlAs$_x$Sb$_{1-x}$ 1µm |
| AlSb 0.1µm |
| GaAs 0.1µm |
| SI GaAs substrate |

→ Superlattice of AlSb & AlAS
666 periods

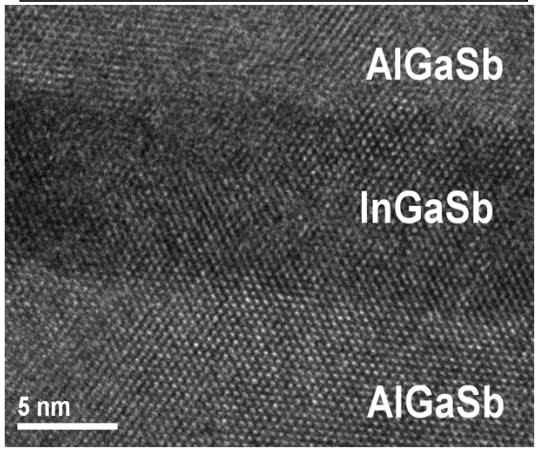
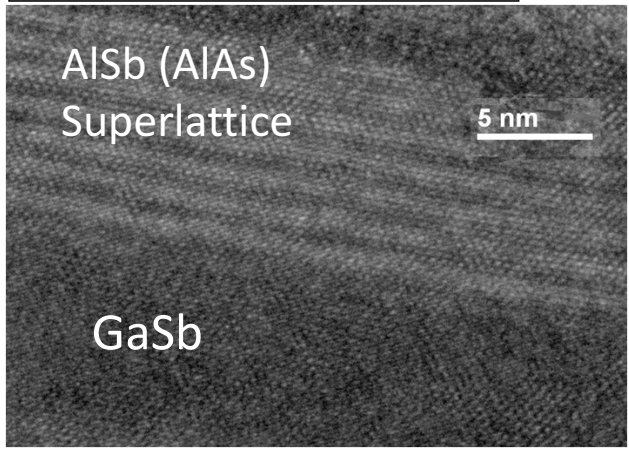

Figure 8 : Cross-section showing the different layers in a quantum-well heterostructure with (A) In$_X$Ga$_{1-X}$Sb and (B) GaSb channel. The AlAs$_X$Sb$_{1-X}$ layer is composed of AlSb/AlAs short-period superlattice. Also shown are high resolution TEM images around the channel region.

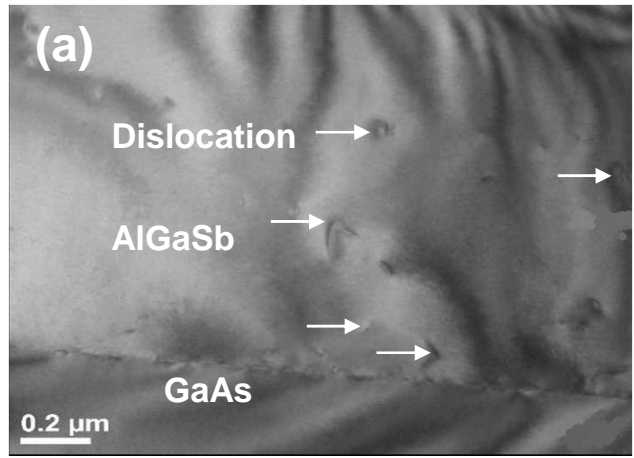

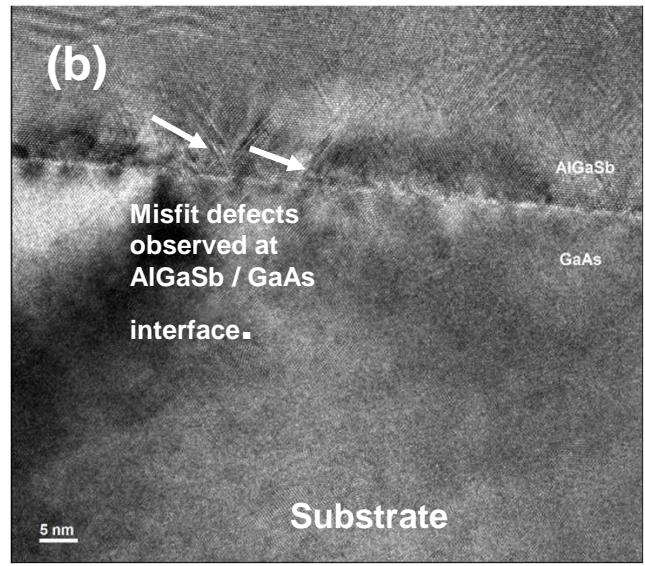

Figure 9: (a) Dislocations and (b) misfit defects in the buffer layer which accommodates the large lattice mismatch between the channel and the GaAs substrate.

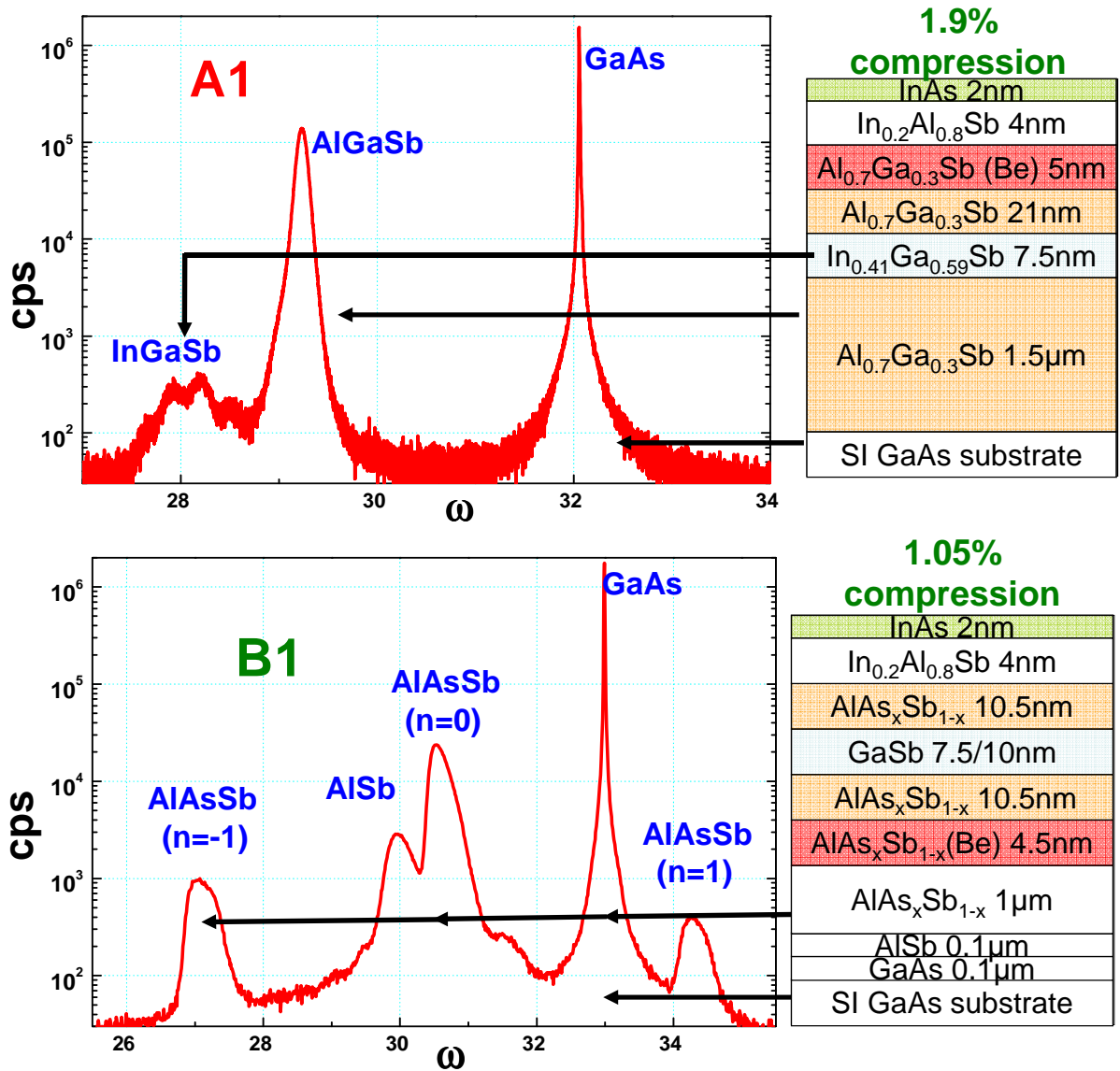

Figure 10: High Resolution XRD scans on the samples A1 (top) & B1 (bottom) near (004) GaAs peak. For sample B1 which uses (AlAs)AlSb as the buffer we observe main and satellite peaks characteristic of the digital superlattice.

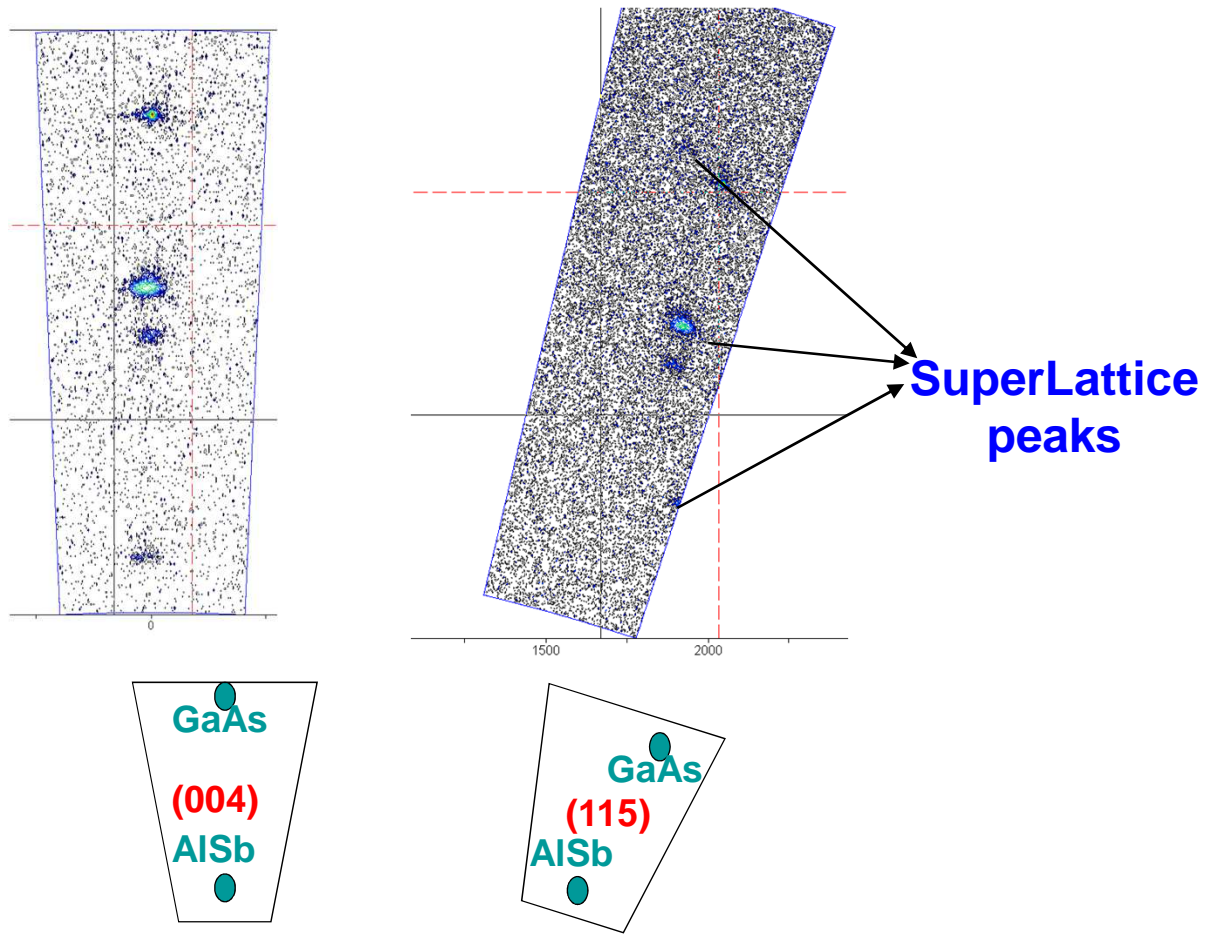

Figure 11: Reciprocal lattice scan on sample B1 around GaAs (004) and (115).

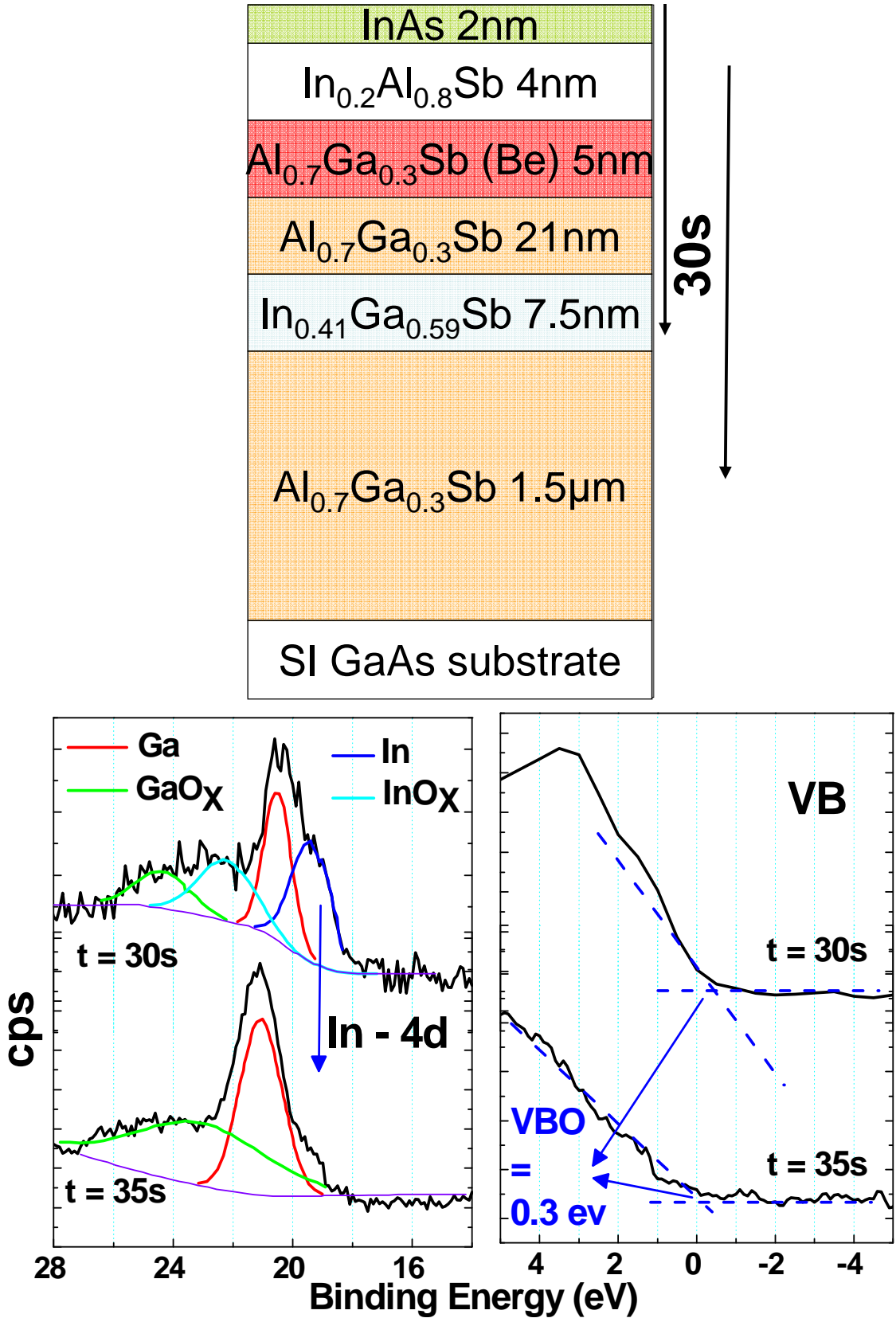

Figure 12: VBO for sample A1 (approach A1) is calculated by taking the difference in the valence band spectrum from the $In_xGa_{1-x}Sb$ channel and $Al_yGa_{1-y}Sb$ buffer.

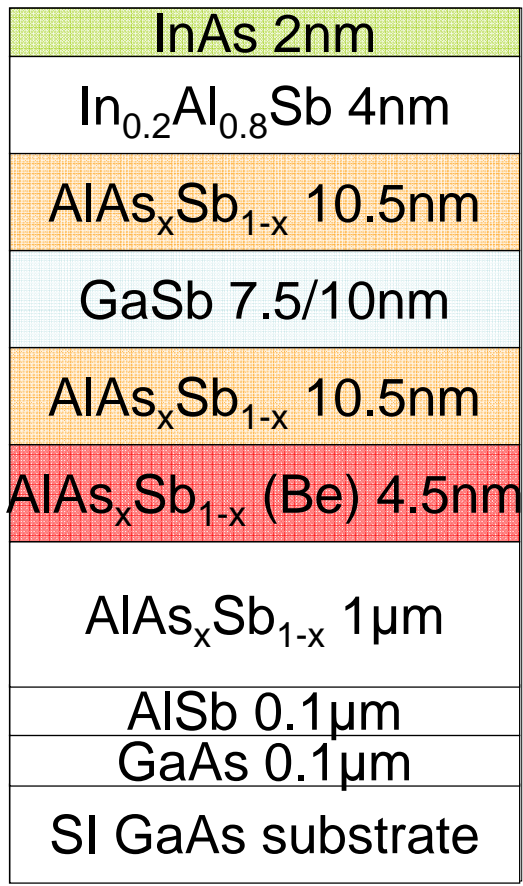
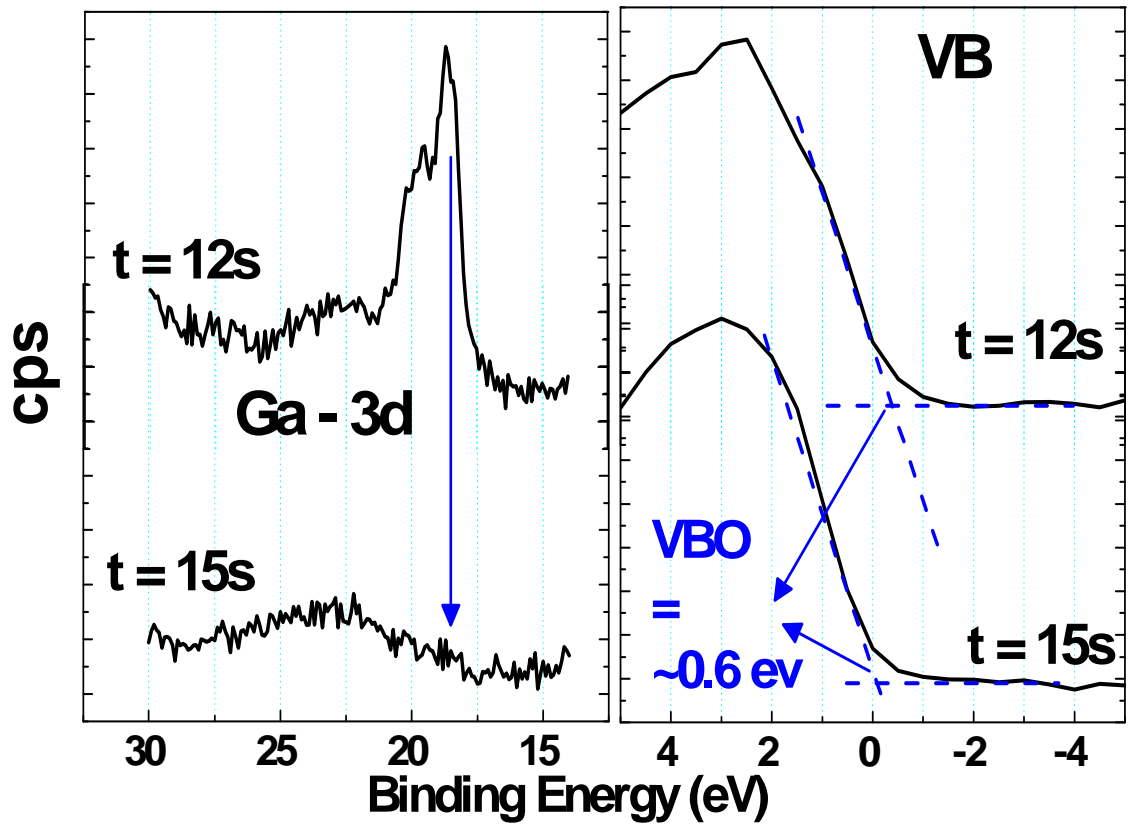

Figure 13: VBO for sample B1 (approach B) is calculated by taking the difference in the valence band spectrum from the GaSb channel and AlAs$_y$Sb$_{1-y}$ buffer.

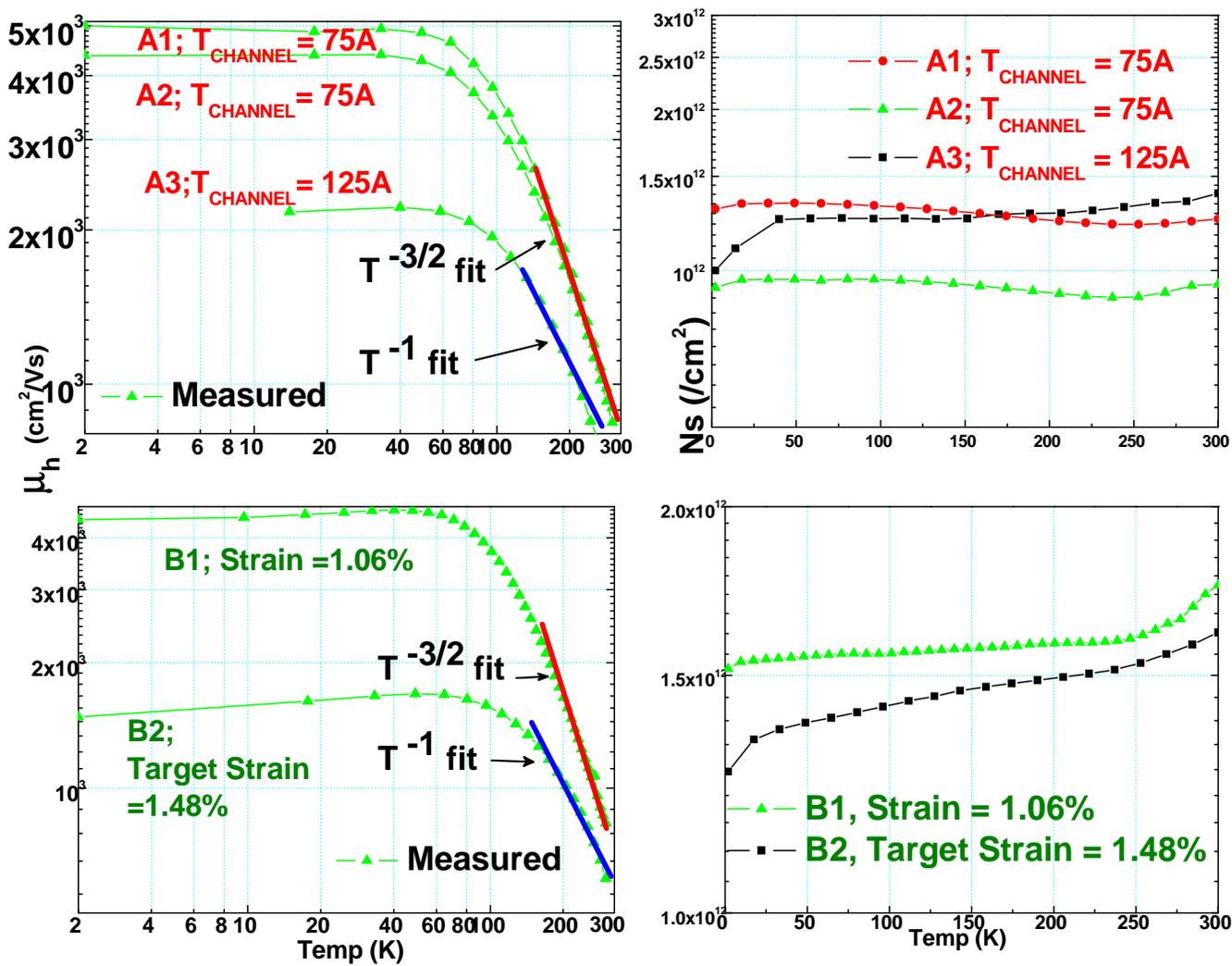

Figure 14 : Hole mobility ($\mu_h$) and sheet charge ($N_s$) are measured as a function of temperature using Hall measurements for samples : A1, A2, A3 (top) and B1, B2 (bottom).

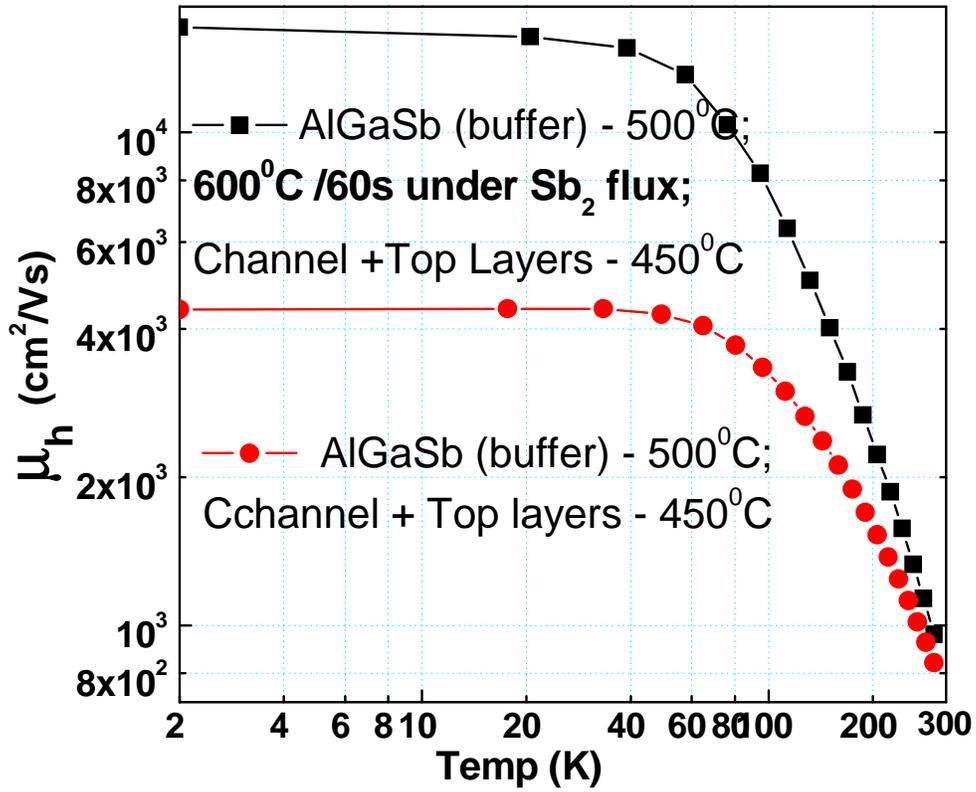

Figure 15: A high temperature anneal (600°C/60s) before channel growth to optimize the interface results in a large increase in low temperature mobility but gives only slight gain (900cm$^2$/Vs to 940cm$^2$/Vs) in mobility at 300K.

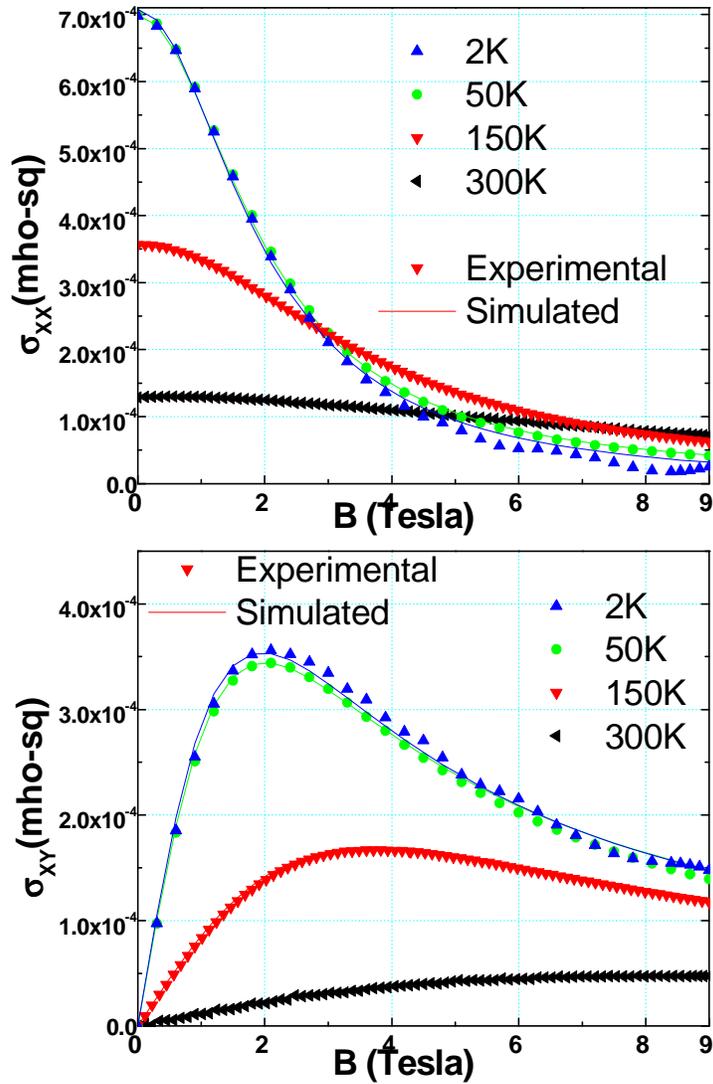

Figure 16: Conductivity tensors ($\sigma_{xx}$ and $\sigma_{xx}$) are measured as a function of magnetic field (B) for various temperature. Mobility spectrum analysis (MSA) on the data confirms that there is no parallel conduction in the stack and is used to estimate number of carriers in *lh*/*hh* bands and their mobility (Figure 17).

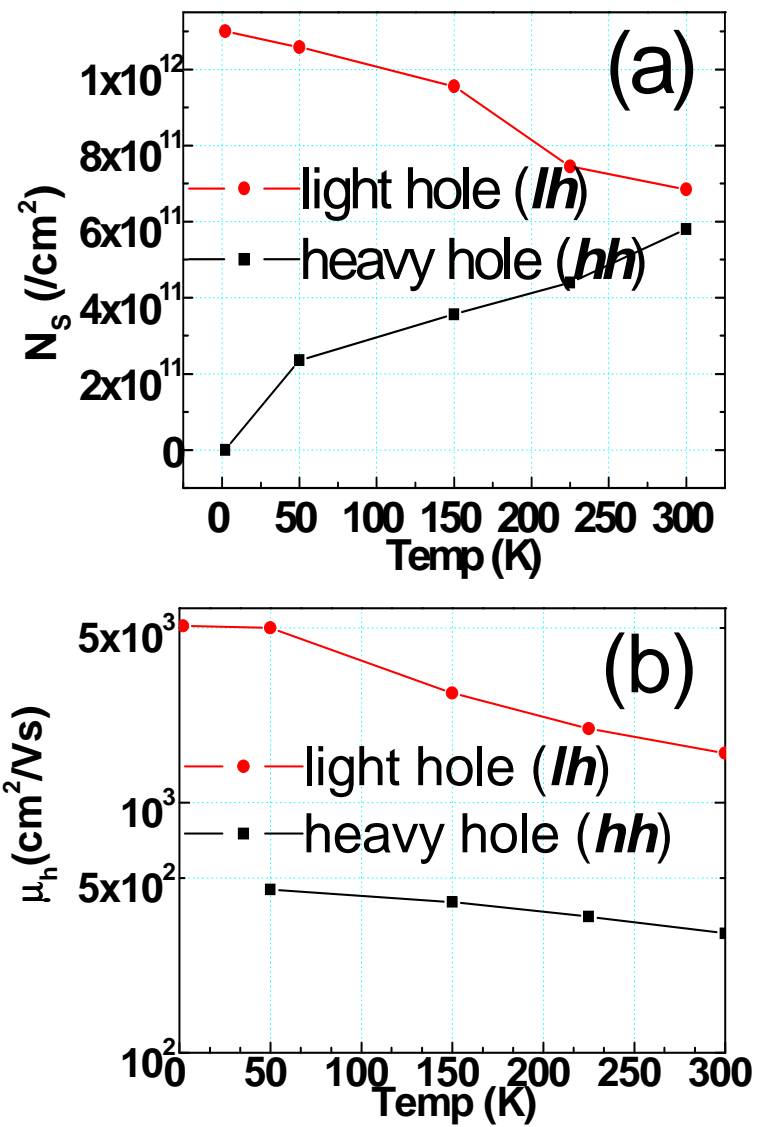

Figure 17: (a) Number and (b) mobility of carriers in the *llight* (*lh*) & heavy hole (*hh*) bands as a function of temperature for sample A1.

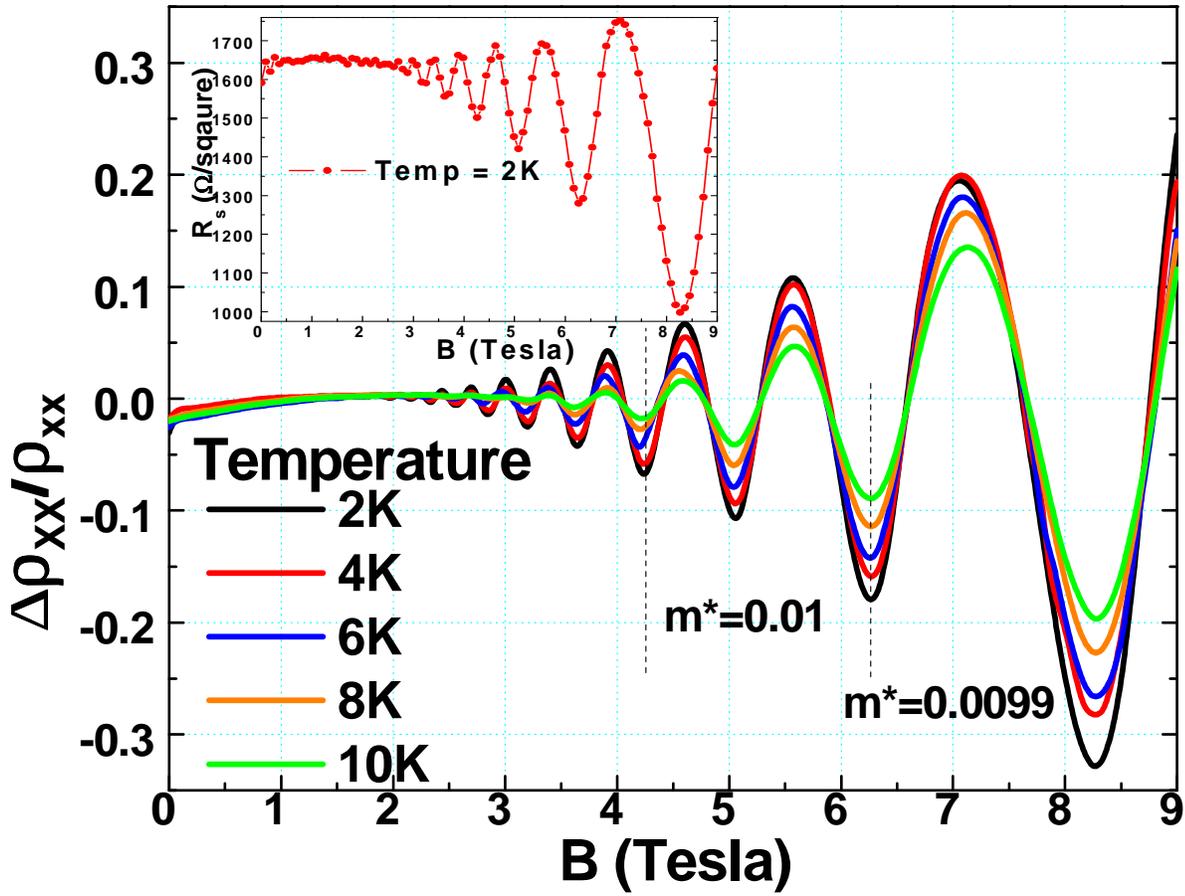

Figure 18: Shubnikov-de-Haas (SdH) oscillations in sheet resistance (inset) are observed at low temperatures and high magnetic field. Temperature dependence of these oscillations is used to calculate **m*** (Table III).

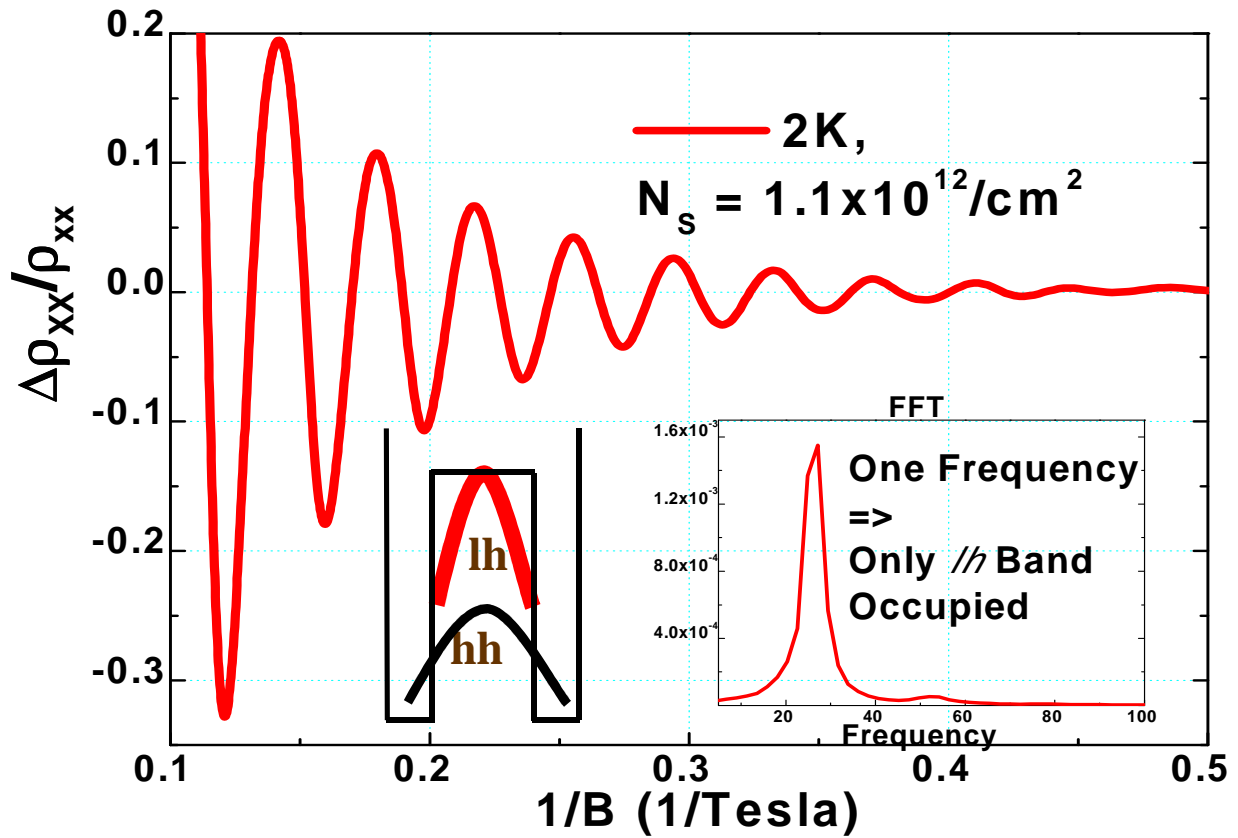

Figure 19 : SdH oscillations at 2K are plotted vs. 1/B for sheet charge of $1.1 \times 10^{12}/cm^2$. The oscillatory behavior is periodic in nature with a single dominant frequency, indicating that only the *lh* band is occupied at this sheet charge.

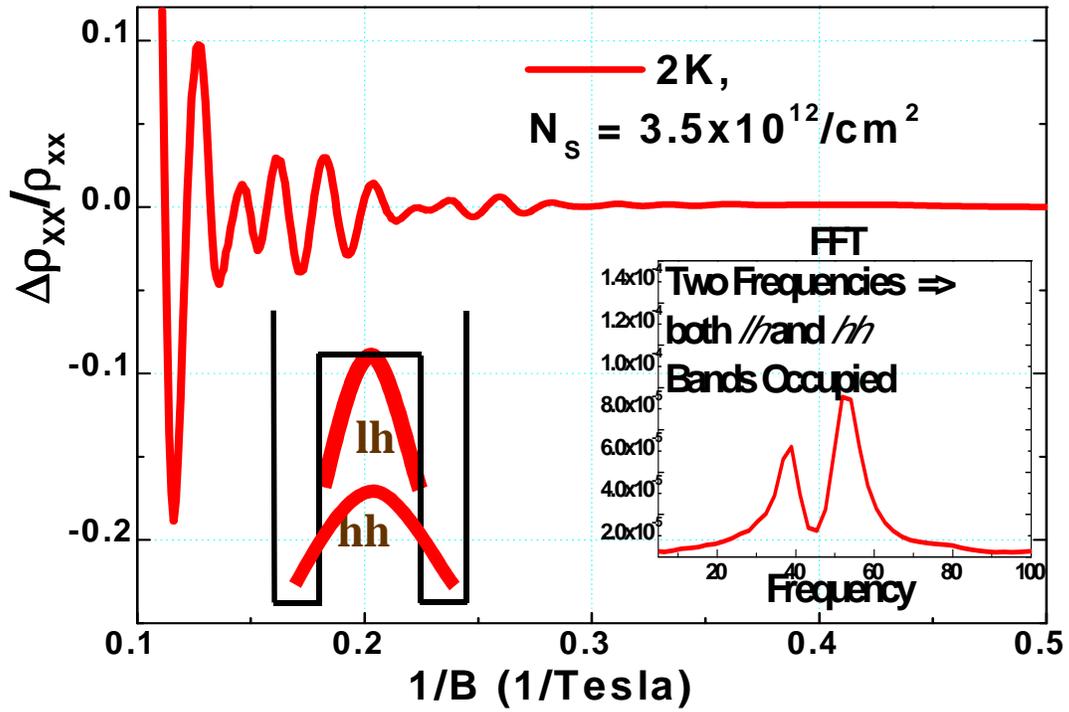

Figure 20 : SdH oscillations at 2K are plotted vs. 1/B for sheet charge of $3.5 \times 10^{12}/cm^2$. The oscillations are combinations of 2 dominant frequencies, indicating that both *lh* and *hh* band is occupied at this sheet charge.

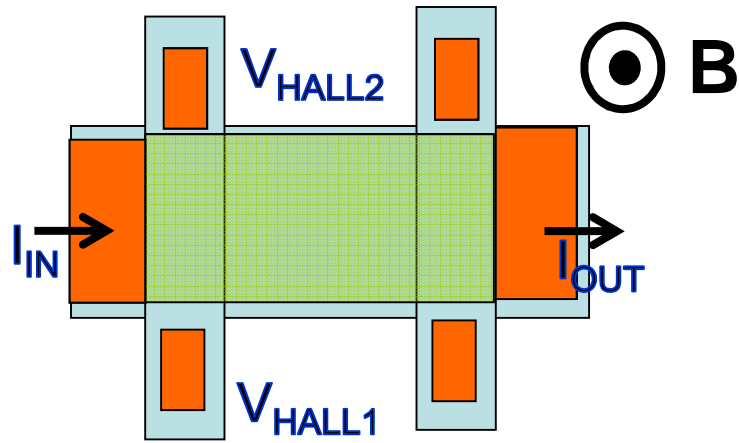
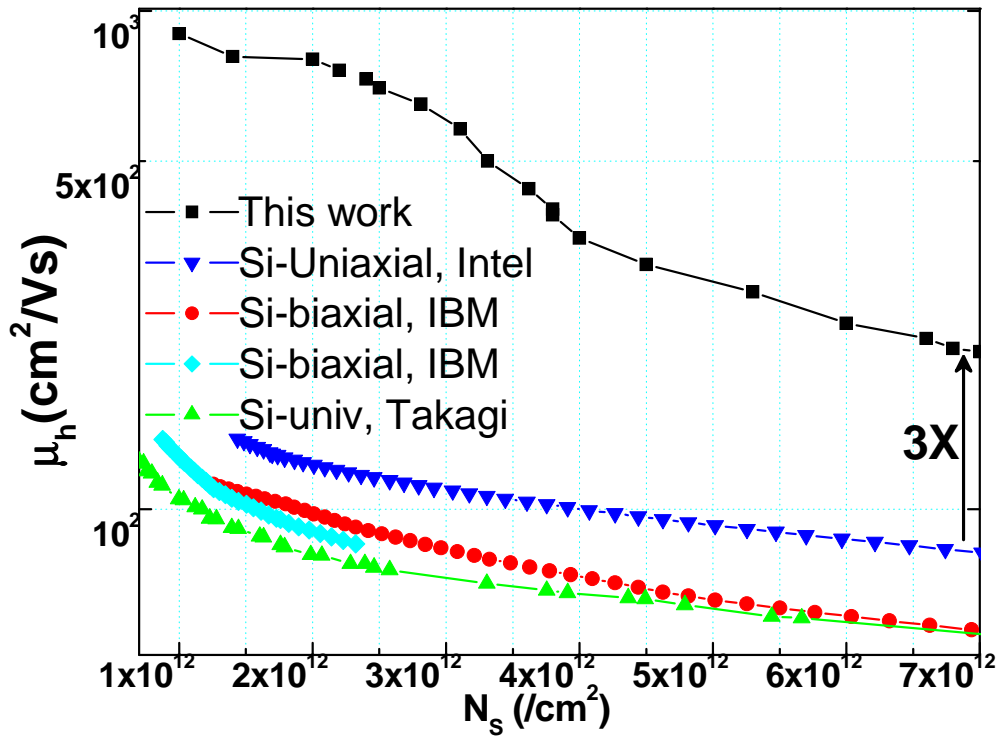

Figure 21 : Hole mobiliyt ($\mu_h$) is measured as a function of sheet charge ($N_s$) using gated hall measurements. Reported values in (strained) silicon are also plotted for comparison.